\documentclass[12pt]{iopart}
\usepackage{amssymb}
\usepackage{harvard}
\usepackage{graphicx}
\usepackage{subfigure}
\usepackage{hyperref}
\let\citet\citeasnoun

\begin{document}
\title[Photodetachment as destruction mechanism for CN$^-$ and C$_3$N$^-$ anions ...]{PHOTODETACHMENT AS DESTRUCTION MECHANISM FOR CN$^-$ and C$_3$N$^-$ ANIONS IN CIRCUMSTELLAR ENVELOPES}
\author{S. S. Kumar$^1$, D. Hauser$^1$, R. Jindra$^1$, T. Best$^1$, \v{S}. Rou\v{c}ka$^2$, W. D. Geppert$^3$, T. J. Millar$^4$, and R. Wester$^1$}

\address{$^1$Institut f\"{u}r Ionenphysik und Angewandte Physik, Universit\"{a}t Innsbruck, A$-$6020 Innsbruck, Austria}
\address{$^2$Charles University in Prague, Faculty of Mathematics and Physics, Department of Surface and Plasma Science, 18000 Prague, Czech Republic}
\address{$^3$Department of Physics, AlbaNova, Stockholm University, SE$-$10691 Stockholm, Sweden}
\address{$^4$Astrophysics Research Centre, School of Mathematics and Physics, Queen's University Belfast, Belfast BT7 1NN, UK}
\ead{roland.wester@uibk.ac.at}
\begin{abstract}
	Absolute photodetachment cross sections of two anions of astrophysical importance CN$^-$ and C$_3$N$^-$ were measured to be (1.18 $\pm$ (0.03)$_{\rm stat}$(0.17)$_{\rm sys}$) $\times$ 10$^{-17}$ cm$^2$ and (1.43 $\pm$ (0.14)$_{\rm stat}$(0.37)$_{\rm sys}$) $\times$ 10$^{-17}$ cm$^2$ respectively at the ultraviolet wavelength of 266 nm (4.66 eV). These relatively large values of the cross sections imply that photodetachment can play a major role in the destruction mechanisms of these anions particularly in photon-dominated regions. We have therefore carried out model calculations using the newly measured cross sections to investigate the abundance of these molecular anions in the cirumstellar envelope of the carbon-rich star IRC+10216. The model predicts the relative importance of the various mechanisms of formation and destruction of these species in different regions of the envelope. UV photodetachment was found to be the major destruction mechanism for both CN$^-$ and C$_3$N$^-$ anions in those regions of the envelope, where they occur in peak abundance. It was also found that photodetachment plays a crucial role in the degradation of these anions throughout the circumstellar envelope.
\end{abstract}
\maketitle
\section{Introduction}

\subsection{Molecular anions in space: Observation and astrophysical relevance}
The discovery of molecules in the interstellar medium about seven decades ago was particularly intriguing since the chemistry governing the formation of molecules in such hostile regions of space was not familiar. Gradually the number of molecules and their cations found in extraterrestrial space increased and it became clear that we in fact live in a ``molecular Universe" \cite{Larsson2012}. However, after the first molecule was identified in the interstellar medium, it took six decades before a molecular anion could be discovered in such environment. This delay was primarily due to the low abundance of anions in space compared to their neutral counterparts and due to the lack of laboratory measurements of high resolution rotational spectra of the anions that could allow their search in space. The first molecular anion ever observed outside our solar system is C$_6$H$^-$ \cite{McCarthy2006}, which was detected in the envelope of the carbon-rich star IRC+10216. The identification of this molecular anion was followed by the discovery of several other carbon chain anions, C$_n$H$^-$ ($n=$ 4, 8) and C$_n$N$^-$ ($n=$ 1, 3, 5) in various regions of space such as dark clouds, circumstellar envelopes, and also in Titan's atmosphere \cite{Cernicharo2007,Cernicharo2008,Sakai2007,Sakai2008,Sakai2010,Agundez2008,Agundez2010,Remijan2007,Brunken2007,Kasai2007,Kawaguchi2007,Thaddeus2008,Gupta2009}. Of these anions, C$_5$N$^-$ was only tentatively identified. The role of anions in the synthesis of molecules in the interstellar medium has been investigated by \citet{Dalgarno1973} many years ago, whereas the formation of molecular hydrogen in stars from H$^-$ had been pointed out by \citet{McDowell1961} much earlier.  The recent discovery of anions in extraterrestrial environments has initiated a fresh interest towards the understanding of anion chemistry in exotic environments. The importance of gas-phase molecular ions in space has been described in detail in a recent review by \citet{Larsson2012}.

\subsection{Significance of the present work}
Extraterrestrial molecular anions are believed to be produced predominantly via electron capture processes such as dissociative or radiative attachment \cite{Larsson2012}. The destruction processes are largely due to photodetachment, associative detachment and mutual neutralization reactions. In photon-dominated regions, the abundance of molecular anions can be mainly determined by their UV photodetachment. Even in the dark clouds where UV photons cannot penetrate UV photodetachment may still contribute to photodestruction of anions because the secondary electrons produced by cosmic rays can excite the molecules to high Rydberg states, which emit UV radiation upon decay. The study of photodetachment processes is also of particular importance in fundamental physics since the extra electron in an anion is bound to the system by means of strong correlated motion of the electrons in the system and the electron-electron correlation plays the most crucial role in such processes. In addition, no theoretical or experimental values of photodetachment cross sections for CN$^-$ or C$_3$N$^-$ have been reported in the literature despite the fact that there have been a number of studies on these molecular anions \cite{Andersen2001,Bradforth1993,Gottlieb2007,Yen2010}. Since these anions have high electron affinities (CN$^-$: 3.862 $\pm$ 0.004 eV \cite{Bradforth1993}, C$_3$N$^-$: 4.305 $\pm$ 0.001 eV \cite{Yen2010}) their photodetachment requires photons in the ultraviolet range. In the present work, we measured the photodetachment cross sections of CN$^-$ and C$_3$N$^-$ anions at an energy (4.66 eV) near the photodetachment thresholds. Furthermore, we used the measured values as an input to model calculations to investigate the impact of the new cross sections on the predicted abundance of anions in the circumstellar envelope of IRC+10216.

\section{Experimental Method and Theoretical Modelling}

\subsection{Experimental Setup}
The basic elements of the experimental setup are an ion source, an octupole ion trap, an MCP detector and a laser system. The ion source consists of a piezoelectric pulsed gas valve with a pair of electrodes (referred to as `plasma electrodes') attached at the exit of the valve. A suitable gas mixture is sent through the gas valve at a certain repetition rate. The ions are generated in a pulsed DC discharge of the gas jet between the plasma electrodes when a high potential difference is applied between them. These ions are then extracted towards the ion trap by a Wiley-McLaren time-of-flight spectrometer oriented perpendicular to the gas jet from the piezo valve. Deflection plates and lenses are used for guiding and focusing the ions into the ion trap. The unique octupole ion trap is made of 100 $\mu$m gold plated molybdenum wires unlike in conventional designs where rods of specific diameter are used as RF electrodes. A short description of this ion trap has been provided by \citet{Deiglmayr2012}, where it was used in conjunction with a magneto-optical trap to study reactive collisions of trapped OH$^-$ anions with trapped rubidium atoms.  A second piezoelectric pulsed gas valve allows for helium buffer gas cooling of the trapped ions. There are two additional electrodes (termed as `shield plates') above and below the trap that enable us to shape the ion density distribution inside the ion trap. The use of thin wires to construct the trap allows one to probe the trapped ions, for instance with a laser, from the sides. The laser beam can be focused at various positions inside the trap by means of a two-dimensional translation stage with a lens attached to it. This configuration is used to map the ion density distribution inside the trap.

In the present experiments, CN$^-$ and C$_3$N$^-$ anions were generated by passing argon gas (at a pressure of about 2--3 bar) over acetonitrile vapor and sending the resulting mixture into the source piezo valve which was operated at 14 Hz. The discharge between the plasma electrodes ionized the gas mixture resulting in the production of several anions including CN$^-$ and C$_3$N$^-$. The plasma was stabilized by the electrons emitted from a hot filament placed opposite to the pulsed gas valve. The ions were injected into the Wiley-McLaren region, where they were extracted towards the trap with an average kinetic energy of about 240 eV. The desired ionic species can be stored in the ion trap by appropriate timing of the switchable voltages applied on the entrance and exit electrodes of the ion trap in accordance with the time of flight of the various molecular ions. The trap was operated at a radiofrequency of 9 MHz with an amplitude of 180 V on top of a DC voltage of about 240 V. The DC voltage of the trap served to reduce the kinetic energy of the ions coming from the source region to about a few eV. The entrance and exit endcap electrodes of the ion trap were between 10 V and 30 V, the exact value of which did not significantly affect the ion distribution except that the signal strength was slightly modified. 

The photodetachment measurements were performed with a pulsed laser beam (266 nm, 10 Hz) obtained by frequency quadrupling of the output from a 1064 nm IR laser system with output pulse energy of about 30 mJ and with pulse width of 7 ns. The pulse energy of the laser beam was reduced to a few tens of microjoule and was then sent through a beam splitter. The transmitted beam was used to measure the fluctuations in the laser energy throughout the experiment and these data were used to correct the measured photodetachment cross section. The reflected beam was focused into the trap using the lens attached on the translation stage. The pulse energy of the beam fired into the trap was as low as 25 $\mu$J so as to ensure that there was only single photon absorption and that the wires constituting the trap were not damaged when the laser beam struck them.

\subsection{Measurement procedure}
The measurement procedure was very similar to the one described previously \cite{Trippel2006,Hlavenka2009,Best2011} except that in the present experiments the laser beam was sent into the ion trap perpendicular to its symmetry axis. Briefly, the ions can be stored in the trap for a few hundred seconds ($1/e$ lifetime, determined from the exponential decay of the ions stored in the trap). In the first part of the experiment, the background decay rate was determined by measuring the amount of ions left in the trap after different storage times. In the second part, the rates of decay were measured with the UV laser pointing at different positions inside the trap. The photodetachment decay rates when plotted as a function of the positions form a tomography image which reflects the ion density distribution inside the trap. The integral of the rate map is proportional to the photodetachment cross section as detailed elsewhere \cite{Trippel2006,Best2011}.

The photodetachment cross section, $\sigma_{\rm pd}$, is given by the expression \cite{Trippel2006,Best2011}:
\begin{equation}\label{Eq:Sigma_expt}
\sigma_{\rm pd} = \frac{1}{\Phi_L}\int[{k_{\rm pd}(x,y)-k_{\rm bg}]\ dx\ dy},
\end{equation}
where $k_{\rm pd}(x,y)$ is the position dependent decay rate due to photodetachment, $k_{\rm bg}$ is the background decay rate (decay rate measured without laser) and $\Phi_L$ is the photon flux.

\subsection{Model Calculations for IRC+10216}
The photodetachment cross sections measured for CN$^-$ and C$_3$N$^-$ anions in the present experiments, together with those obtained previously \cite{Best2011} for the carbon chain anions C$_2$H$^-$, C$_4$H$^-$ and C$_6$H$^-$, were used as input for model calculations of the circumstellar envelope of IRC+10216. The photodetachment cross sections of CN$^-$ and C$_3$N$^-$ were fitted to the expression used in previous model calculations \cite{Millar2007}:
\begin{equation}\label{Eq:Sigma_model}
	\sigma = \sigma_{\infty}\sqrt{1-E_{\rm A}/\epsilon},
\end{equation}
in which $\sigma$ is the cross section, $\sigma_\infty$ the cross section at infinite photon energy, $E_{\rm A}$ the photodetachment threshold energy and $\epsilon$ the photon energy. Since our cross section measurements have been carried out only at a wavelength of 266 nm (at which a sufficiently intense UV beam was available from our laser systems) only two points exist for the fit of the photon energy/cross section curve for each of the nitrile anions, the cross section at threshold energy (where $\sigma = 0$) and the one measured at 266 nm. Of course, there exists the possibility of strong resonances, especially at photon energies only slightly above the threshold, which cannot be ruled out in the absence of complementary experimental data on photodetachment cross sections of these anions. For a more accurate treatment of the model, one would require experimental cross sections at higher photon energies necessitating radiation from sources such as synchrotrons or free electron lasers (FELs). Regarding the threshold energy ($E_{\rm A}$) of CN$^-$, the value obtained by \citet{Bradforth1993} who employed a pulsed fixed-frequency negative ion photoelectron spectrometer (3.862 $\pm$ 0.004 eV) was used for the fit. In the case of C$_3$N$^-$ the result from \citet{Yen2010} measured using slow electron velocity-map imaging (4.305 $\pm$ 0.001 eV) and field-free time-of-flight was applied. For the photodetachment cross sections of the hydrocarbon anions the fitted values from \citet{Best2011} were used.

The chemical models are based on the assumption of a uniform mass-loss rate for the circumstellar envelope of IRC+10216 described by \citet{Millar2000} together with a second model, described by \citet{Cordiner2009}, in which density-enhanced shells are included. With density-enhanced shells of gas and dust, a more realistic modelling of the circumstellar envelope is achieved by introducing a set of density enhancements with the physical parameters of the envelope based on the dust-shell observation by \citet{Mauron2000}.  For modelling, the conditions expected for the well-studied circumstellar envelope of IRC+10216 were applied. Consequently, a spherically symmetric outflow velocity from the central star of 1.45 $\times$ 10$^6$ cm s$^{-1}$ and a mass loss of 1.5 $\times$ 10$^{-5}$ solar masses per year were assumed for the envelope \cite{Menshchikov2001}. The adopted temperature profile is based on a fit to the gas kinetic temperature profile of \citet{Crosas1997}, with a minimum temperature of 10 K fixed in the outer region of the envelope and is the same as used by \citet{Cordiner2009}. The initial chemical abundances of parent molecules relative to that of H$_2$ used in the model are listed in Table \ref{T:Abund}. These species are formed in the inner envelope close to the star at high density and temperature and blown outwards in a spherically symmetric outflow \cite{Millar2000}. The calculations begin at an inner radius of $10^{15}$ cm where photons from the external, interstellar radiation field begin to destroy parent species creating reactive radicals and ions and initiating the synthesis of anions and other species. The number density $n(r)$ declines with the radius as $1/r^2$. In the second set of calculations, in addition to the $1/r^2$ dependence of the number density, a series of step-like density enhancements of the form $\beta n(r)$ is introduced. The parameter $\beta$ is set to 5 for all shells in the model. According to dust shell parameters deduced from scattered light observations by \citet{Mauron2000}, we assume that each shell has a thickness of 2 arcsec and the spacing between the shells is 12 arcsec. This distance corresponds to roughly 530 years between the peaks of enhanced mass loss (See \citet{Cordiner2009} for details.).
\begin{table}
	\centering
	\begin{tabular}{cc}
		\hline
Parent molecule & Abundance relative to H$_2$ \\
		\hline
CS    & 7.0 $\times$ 10$^{-7}$ \\
SiO   & 1.8 $\times$ 10$^{-7}$ \\
SiS   & 1.3 $\times$ 10$^{-6}$ \\
CO    & 6.0 $\times$ 10$^{-4}$ \\
C$_2$H$_2$  & 8.0 $\times$ 10$^{-5}$ \\  
HCN   & 2.0 $\times$ 10$^{-5}$ \\  
CH$_4$& 3.5 $\times$ 10$^{-6}$ \\
NH$_3$& 2.0 $\times$ 10$^{-6}$ \\
SiH$_4$& 2.2 $\times$ 10$^{-7}$ \\
SiC$_2$& 2.0 $\times$ 10$^{-7}$ \\
H$_2$O& 1.0 $\times$ 10$^{-7}$ \\
HCl& 1.0 $\times$ 10$^{-7}$ \\
HCP   & 2.5 $\times$ 10$^{-8}$ \\
C$_2$H$_4$  & 2.0 $\times$ 10$^{-8}$ \\
HF & 8.0 $\times$ 10$^{-9}$ \\
H$_2$S & 4.0 $\times$ 10$^{-9}$ \\
N$_2$  & 2.0 $\times$ 10$^{-4}$ \\
Mg    & 1.0 $\times$ 10$^{-5}$ \\      
He & 1.0 $\times$ 10$^{-1}$ \\
		\hline
	\end{tabular}
	\caption{Abundances of parent species used in the model.}\label{T:Abund}
\end{table}

\section{Results and Discussion}

\subsection{Experimental results}
\begin{figure}[h]
	\centering
	\mbox{\subfigure{\includegraphics[scale=0.39,clip]{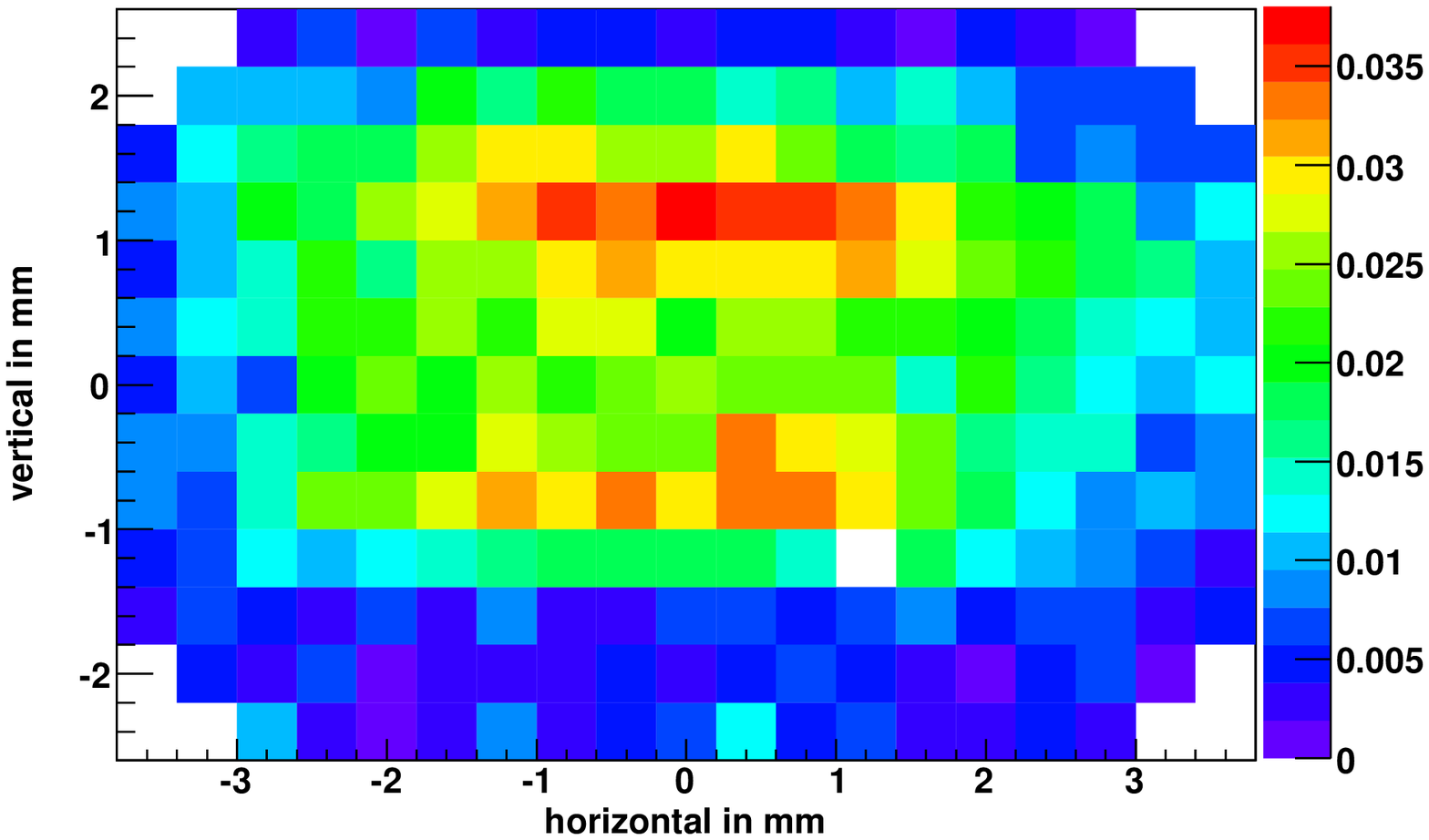}}
	\subfigure{\includegraphics[scale=0.39,clip]{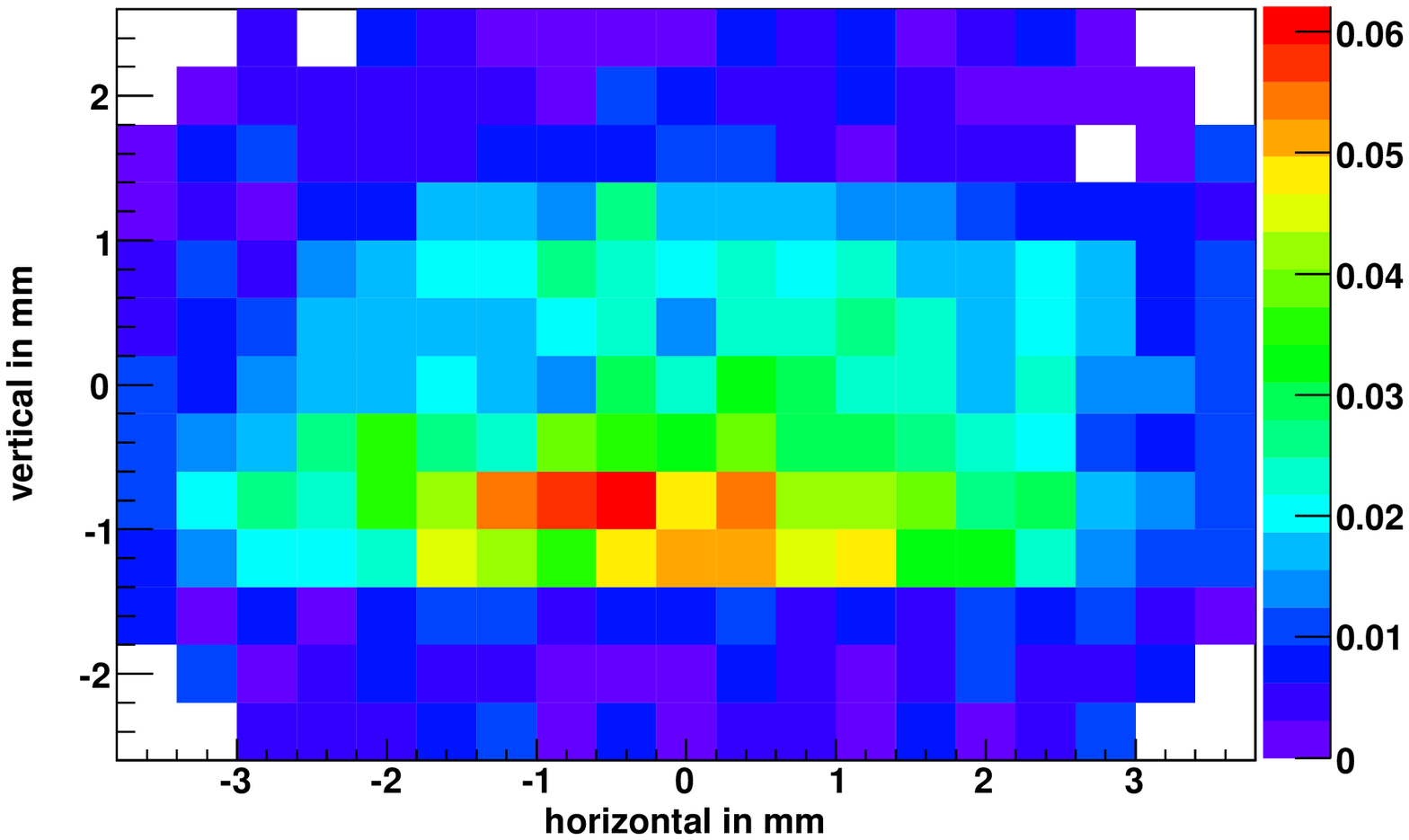} }}
	\caption{Tomography images for the CN$^-$ ions for two different ion density distributions in the trap. The numbers on the color bars are in the units of s$^{-1}$. The cross sections determined from these distributions agree to within 4\%. Similar results were obtained for C$_3$N$^-$ also (not shown).} \label{F:CN}
\end{figure}

Figure \ref{F:CN} presents the tomography images for the CN$^-$ anions at two different configurations of the shield plate voltages. One can clearly see a difference in the ion distributions inside the trap. In fact, the ion trap exhibits two local minima in the vertical direction due to the presence of the holes in the shield plates placed above and below the trap. By adjusting the voltages on these plates, one can redistribute the ions in the trap into these local minima. On the left-hand side of Figure 1, the ions are more or less equally distributed in the two minima, whereas on the right-hand side, the lower minimum is mostly populated. The cross sections measured from these strongly different distributions agree to within 4\%. A similar procedure was employed for C$_3$N$^-$. For C$_3$N$^-$, the error is larger, about 10\%, because its signal strength was almost an order of magnitude less than that of CN$^-$, and hence the fluctuations in the ion signal limited the accuracy with which the rates could be determined. The values of the measured cross sections for both CN$^-$ and C$_3$N$^-$ from different measurements are summarized in Table \ref{T:CNC3N}. The determination of the systematic uncertainties (as percentage error) in the measurements involves several factors which are listed in Table \ref{T:Err}.

\begin{table}[h]
	\centering
	\begin{tabular}{ccccccc}
		\hline
	No. of measurements $\rightarrow$ & 1 & 2 & 3 & Average & $\sigma_\infty$ \\
	\hline
	CN$^-$ & 1.14 & 1.18 & 1.21 & 1.18 & 2.84\\
       C$_3$N$^-$ & 1.32 & 1.39 & 1.58 & 1.43 & 5.19\\
	\hline
	\end{tabular}
	\caption{Cross sections ($\times 10^{-17}$ cm$^2$) from a few sets of measurements on CN$^-$ and C$_3$N$^-$ at 266 nm together with the average values and the values from the fit to Equation (\ref{Eq:Sigma_model}). For accuracy, see Table \ref{T:Err}.}\label{T:CNC3N}
\end{table}

\begin{table}[h]
        \centering\footnotesize
        \begin{tabular}{ccccc}
		\hline
	Source of Error   & Error (CN$^-$)  & Max. Error (CN$^-$) & Error (C$_3$N$^-$) & Max. Error (C$_3$N$^-$)\\
	\hline
	Laser energy fluctuation & 7.6 & 10 & 12.3 & 15\\
   	Absorption of laser by window & 0.5 & 1 & 0.5 & 1\\
   	Reflection coeff. of beam splitter & 1.0 & 2 & 1.0 & 2 \\
	Imaging aspect ratio & 0.8 & 2 & 0.8 & 2\\
	Integration limits for ion signal & 0.9 & 2 & 3.9 & 4 \\
   	Background subtraction & 2.2 & 3 & 5.9 & 8 \\
	Overlap of laser beam with wires & 1.3 & 1.5 & 1.3 & 1.5 \\
	\hline
	Total & 14.3 & 21.5 & 25.7 & 33.5\\
	\hline
	\end{tabular}
	\caption{Possible contribution of errors (\%) in calculating the cross sections of CN$^-$ and C$_3$N$^-$. The various contributions are assumed to be independent. Integration limits and background subtraction are in fact not completely independent. However, the dependence is not systematic. Further, this correction does not make any significant difference in the estimation of errors.}\label{T:Err}
	\end{table}

The photodetachment cross sections of CN$^-$ and C$_3$N$^-$ are a factor of at least two larger than the cross sections measured for other carbon chain anions C$_n$H$^-$ \cite{Best2011}. Hence the abundance of these cyano anions in photon-dominated regions is likely to be significantly influenced by their photodetachment. Furthermore, the cross sections are determined at an energy which is not far away from the photodetachment threshold for both CN$^-$ and C$_3$N$^-$. Therefore, the large cross section values may indicate the presence of strong resonances.

\subsection{Results from model calculations}

\begin{figure}[h]
	\centering
	\includegraphics[scale=0.72,clip]{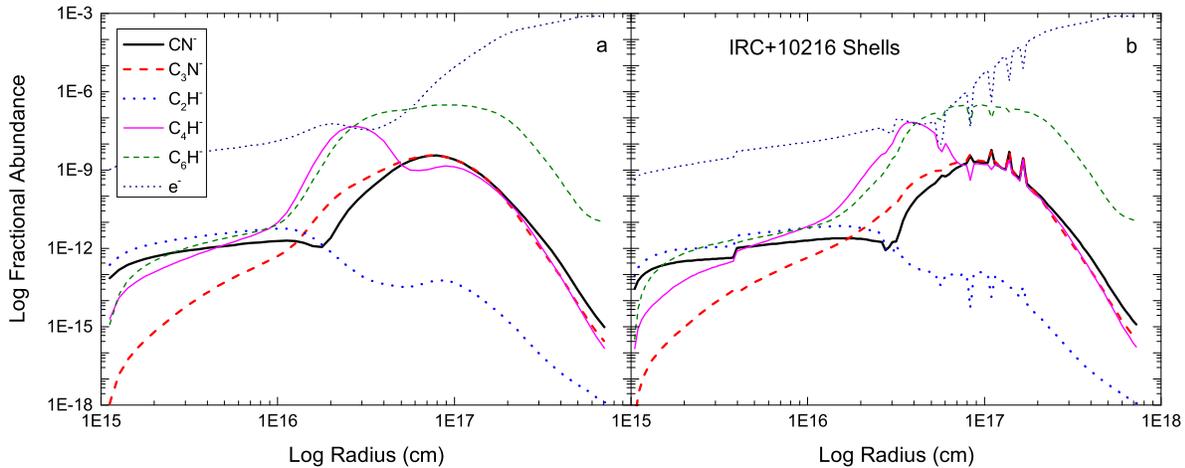}
	\caption{Fractional abundances of the important anions, as well as the electron fraction, as a function of radius in the circumstellar envelope of IRC+10216 without (a) and with (b) shells.} \label{F:Rate12}
\end{figure}

 The photodetachment rate constants that function as input data for the model calculations were obtained using a standard interstellar radiation field \cite{Draine1978}. The obtained values were 2.55 $\times$ 10$^{-9}$, 1.99 $\times$ 10$^{-9}$, 1.16 $\times$ 10$^{-9}$, 6.72 $\times$ 10$^{-9}$ and 1.03 $\times$ 10$^{-8}$ s$^{-1}$ for C$_2$H$^-$, C$_4$H$^-$, C$_6$H$^-$, CN$^-$ and C$_3$N$^-$, respectively. In this calculation we have used the entire reaction set and the rate coefficients of the UMIST database for astrochemistry 2012 \cite{McElroy2013}, which includes the additional anion production mechanisms mentioned by \citet{Cordiner2008}, to calculate molecular abundances as a function of the radial distance from the centre of the star (See \citet{Cordiner2009} for details.). In a second set of calculations, shells of matter with densities that are enhanced relative to the surrounding circumstellar medium were included in the model. Figures \ref{F:Rate12}a (no shells) and \ref{F:Rate12}b (with shells) show the fractional abundances of the important anions, as well as the electron fraction, as a function of radius in the circumstellar envelope. When integrated over radius, these abundances yield the total column densities which were compared with those using the cross section function, $\sigma = 1 \times 10^{-17} \sqrt{1-E_{\rm A}/\epsilon}$  cm$^2$, employed in previous studies \cite{Millar2007}. The column densities using these two approaches are listed in Table \ref{T:Coldens}.

\begin{table}[h]
\tiny	\centering
	\begin{tabular}{cccccc}
		\hline
		Anion & Using $\sigma$ from & Using $\sigma$ from & Using experimentally & Using experimentally & Observed column density \\
		      & \citet{Millar2007} & \citet{Millar2007}. & determined $\sigma$ & determined $\sigma$. &  \\
		      & & Model with shells & & Model with shells & \\
		\hline
		CN$^-$ & $6.9 \times 10^{11}$ & $8.3 \times 10^{11}$  & $5.5 \times 10^{11}$ & $7.0 \times 10^{11}$  & $5 \times 10^{12}$\cite{Agundez2010} \\
		C$_3$N$^-$ & $1.0 \times 10^{12}$ & $1.2 \times 10^{12}$  & $6.9 \times 10^{11}$ & $8.7 \times 10^{11}$  & $1.6\pm 0.6 \times 10^{12}$\cite{Thaddeus2008} \\
		\hline
	\end{tabular}\caption{Calculated and observed column densities (in cm$^{-2}$) of anions in IRC+10216.}\label{T:Coldens}
\end{table}

It can be seen that the input of the experimental cross sections somewhat reduces the column densities and thus slightly deteriorates the agreement between the modelled and observed cross sections. Also, the C$_3$N$^-$/C$_3$N ratio predicted by the model ($1.4\times10^{-3}$) now lies below the observed value ($5\times10^{-3}$), whereas previous models tended to overestimate it \cite[and references therein]{Herbst2009}. The inclusion of high-density shells does increase the anion column densities by around 10--20\%. However, they remain smaller, but within the same order of magnitude, than those observed.

The predicted densities not only depend strongly on the rates of photodetachment but also on the efficiency of the formation reactions, such as radiative attachment, radical-ion and dissociative attachment reactions. Regarding the generation of the two cyano anions the model predicts that reactions of N radicals with C$_n^-$ ions, e.g.,
\begin{eqnarray}
	\textrm{C}_6^- + \textrm{N}& \rightarrow &\textrm{C}_3\textrm{N}^- + \textrm{C}_3 \\
	\textrm{C}_6^- + \textrm{N} &\rightarrow &\textrm{CN}^- + \textrm{C}_5
\end{eqnarray}
dominate as formation pathways in the outer and middle parts of the envelope (r $\geq 10^{16}$ cm) for both CN$^-$ and C$_3$N$^-$. These processes might be partly responsible for the extraordinarily high anion to neutral abundance ratio for C$_3$N$^-$ \cite{Cordiner2009,Agundez2010,Thaddeus2008,Cernicharo2007}. In the innermost regions (r $\leq 10^{16}$ cm), formation of CN$^-$ proceeds via reaction of H$^-$ with HCN:
\begin{equation}
\textrm{H}^- + \textrm{HCN} \rightarrow \textrm{CN}^- + \textrm{H}_2.
\end{equation}
The importance of the latter process is due to the formation of H$^-$ through cosmic ray induced ion pair formation in the inner shells of the envelope \cite{Cordiner2009,Prasad1980}. At these small radii, radiative attachment of C$_3$N and dissociative attachment of HNC$_3$ are predominant formation routes of C$_3$N$^-$:

\begin{eqnarray}
	\textrm{C}_3\textrm{N} + \textrm{e}^- & \rightarrow &\textrm{C}_3\textrm{N}^- + \textrm{h}\nu \\
	\textrm{HNC}_3 + \textrm{e}^- &\rightarrow & \textrm{C}_3\textrm{N}^- + \textrm{H}.
\end{eqnarray}
Whereas the reactions of N atoms with C$_n$ chain anions have been characterized in a selected ion flow tube experiment \cite{Eichelberger2007}, there are, as yet, no laboratory studies on the formation of the cyanide anion from H$^-$ and HCN.

There are also uncertainties in the destruction processes. At a distance from the central star of around 6 $\times 10^{16}$ cm, where the abundance of the CN$^-$ and the C$_3$N$^-$ anions peaks, photodetachment clearly is the most important degradation mechanism of the two anions and accounts for 45 \% of the breakdown of C$_3$N$^-$ and 35 \% for CN$^-$. In the case of CN$^-$, other decay processes are mutual neutralization with C$^+$ (30 \%) and Si$^+$ (7 \%) as well as associative detachment with H (15 \%). Minor loss processes of C$_3$N$^-$ are mutual neutralization with C$^+$ (25 \%) and Si$^+$ (6 \%) and associative detachment with H (11 \%). In the outer regions of the cloud (r $> 10^{17}$ cm) mutual neutralization with C$^+$ actually becomes predominant for both C$_3$N$^-$ (accounting for 72 \% of the loss at a distance of $2.5 \times 10^{17}$ cm from the star) and CN$^-$ (79 \% at the same radius). This behavior is most likely due to the increase of C$^+$ abundance towards the edge of the cloud (the peak density of this species there is around 5.4 $\times$ 10$^{-2}$ cm$^{-3}$ with an abundance ratio C$^+$/H$_2$ of $2.1\times10^{-4}$ at a radius of $2.2 \times 10^{17}$ cm), which is caused by photoionization of C through the interstellar radiation field. In the inner regions of the circumstellar envelope (r $ < 10^{16}$ cm) mutual neutralization with Mg$^+$ is predicted to be the main degradation process. This can be explained by the fact that the Mg$^+$ number density is fairly constant throughout the envelope (ranging between $2 \times 10^{-4}$ cm$^{-3}$ and $2 \times 10^{-3}$ cm$^{-3}$), whereas the C$^+$ abundance is as low as $1.5 \times 10^{-6}$ cm$^{-3}$ at a radius of $2 \times 10^{16}$ cm. Consequently, the abundance ratio of C$^+$ to H$_2$ increases from $1.0\times10^{-12}$ at a radius of $2.2\times10^{15}$ cm to $7.8\times10^{-4}$ at a radius of $7.1\times10^{17}$ cm, whereas the one of Mg$^+$ to H$_2$ spans only 5 orders of magnitude, rising from $2.1\times10^{-10}$ at a radius of $2.2\times10^{15}$ cm to $1.0\times10^{-5}$ at a radius of $7.1\times10^{17}$ cm. But even at the outermost and the innermost distances photodetachment significantly contributes to the destruction of CN$^-$ and C$_3$N$^-$.

The peak abundances of the two cyano anions investigated in this study lie at the radii $6.3 \times 10^{16}$ cm and $5.6 \times 10^{16}$ cm for CN$^-$ and C$_3$N$^-$, respectively, and the maxima of the fractional abundances  at  $7.9\times10^{16}$ cm for CN$^-$ and $7.1\times10^{16}$ cm for C$_3$N$^-$. This implies that the CN$^-$ peak radius predicted by the model is somewhat larger than the one concluded from observations ($2 \times 10^{16}$ cm), but slightly lower than the one predicted by model calculations of \citet{Agundez2010} ($8 \times 10^{16}$ cm). From the present data it can be concluded that photodetachment is a very crucial process in the degradation of anions throughout the envelope. However, one has to consider the uncertainties regarding the rate constants of the formation and destruction mechanisms of the two anions. The relative importance of photodetachment depends on the rate constants of the competing processes, namely the mutual neutralization processes of the cyano anions with C$^+$ and other metallic ions. \citet{Harada2008} estimated the rate constant of the reaction of C$_3$N$^-$ with C$^+$ based on earlier flowing afterglow Langmuir probe measurements \cite{Smith1978} of other ions to follow the expression:
\begin{equation}
\textrm{k}=7.5\times10^{-8} (\textrm{T}/300)^{-0.5}\ \textrm{cm}^3\textrm{s}^{-1}.
\end{equation}
To the best of our knowledge, no experimental data on the reaction rate constants of these processes have so far been obtained. The new DESIREE double storage ring at Stockholm University will amend this shortcoming \cite{Schmidt2008}.

In agreement with other model calculations, the abundances of the anions peak at larger radii than the corresponding neutrals \cite{Guelin2011}. The inclusion of shells with enhanced density similar to the model of \citet{Cordiner2009} increases the column densities of the anions by about 20\% and improves the agreement with observed column densities, predominantly through reducing the rates of photodetachment through the increased dust extinction that they provide.

\section{Conclusions}

The absolute photodetachment cross sections of two molecular anions of astrophysical importance, CN$^-$ and C$_3$N$^-$, were measured at the ultraviolet wavelength of 266 nm. The measured cross sections are relatively high and might indicate the possibility of strong resonances near the photodetachment threshold. High cross sections imply that the abundance of these molecular anions can be crucially dependent on their destruction by photodetachment especially in photon-dominated regions. The presented model calculations, carried out to investigate molecular anions in the circumstellar envelope of IRC+10216, predict the relative importance of the various mechanisms of production and destruction of cyano anions in different regions of the envelope. It was found that in regions where these molecular anions have their peak abundance, photodetachment serves as the most important destruction mechanism. The calculations also predict that photodetachment significantly contributes to the destruction of these anions throughout the circumstellar envelope. Thus photodetachment plays a fundamental role in the degradation of anions in circumstellar envelopes. However, its exact significance can only be determined if more data on other competing pathways are available. Future experimental investigations on these processes are therefore vital for our understanding of the anion chemistry of circumstellar envelopes.

This work has been supported by the European Research Council under ERC grant agreement No. 279898 and by the ESF COST Action CM0805 ``The Chemical Cosmos: Understanding Chemistry in Astronomical Environments". We thank Eric Endres for his support during the experiments. Research in molecular astrophysics at QUB is supported by a grant from the STFC. \v S. R. ackowledges support by Czech Grant Agency under contract No. P209/12/0233.

\end{document}